\documentclass[12pt]{article}
\usepackage{amsfonts}

\usepackage{amsmath}
\usepackage{amstext}

\newtheorem{theorem}{Theorem}

\newtheorem{corollary}[theorem]{Corollary}

\newtheorem{lemma}[theorem]{Lemma}

\newtheorem{proposition}[theorem]{Proposition}

\newenvironment{proof}[1][Proof]{\textbf{#1.} }{\ \rule{0.5em}{0.5em}}
\input{tcilatex}

\input{tcilatex}

\begin{document}

\author{Michael Frankel \\
%EndAName
{\small \textit{Department of Mathematical Sciences,}}\\
{\small \textit{Indiana University Purdue University Indianapolis, }}\\
{\small \textit{\ Indianapolis,}} {\small \textit{\ IN 46202-3216 U.S.A.}}
\and Victor Roytburd \\
%EndAName
{\small \textit{Department of Mathematical Sciences, }}\\
{\small \textit{\ Rensselaer Polytechnic Institute, }}\\
{\small \textit{\ Troy, NY 12180-3590, U.S.A.}}}
\title{ON ATTRACTORS FOR A SHARP INTERFACE MODEL OF EXOTHERMIC PHASE TRANSITIONS%
{\normalsize \textbf{\ }}}
\date{\empty}
\maketitle

\begin{abstract}
We study a free interface problem related to combustion of condensed matter
and some non-equilibrium exothermal phase transitions. In spite of a variety
of non-trivial dynamical scenarios exhibited by the model the solutions are
uniformly bounded and the interface velocity is a smooth function. The main
result of the paper establishes existence of a compact connected attractor
for the classical solutions of the problem. Numerical evidence leads to the
conjecture that the fractal dimension of the attractor is finite.

\emph{Submitted to Interfaces and Free Boundaries}
\end{abstract}

% submitted electronically to Interfaces and Free Boundaries on Dec 27,2002

\section{Introduction}

This paper presents a study of attractors for a two-phase Stefan problem
with kinetics. We show that classical solutions of the problem approach a
compact connected attractor in the uniform norm. We also demonstrate via
direct numerical simulations that the attractor has a finite correlation
dimension.

The free-boundary problem that is the subject of the paper\ arises naturally
as a mathematical model of a variety of exothermic phase transition type
processes, such as solid combustion \cite{matsiv-solid} also known as
Self-propagating High-temperature Synthesis or SHS \cite{munir},{\normalsize %
\ } solidification with undercooling \cite{langer}, laser induced
evaporation \cite{gt}, rapid crystallization in thin films \cite{saarloos}
etc. These processes are characterized by production of heat at the
interface, and their dynamics is determined by the feedback mechanism
between the heat release due to the kinetics and the heat dissipation by the
medium. In addition to its theoretical interest SHS has industrial
applications as a method of synthesizing certain technologically advanced
materials for high-temperature semiconductors, nuclear safety devices, fuel
cells etc., see \cite{munir}, \cite{var2} and also \cite{var1} for a popular
exposition. SHS propagates through mixtures of fine elemental reactant
powders (e.g., Ti + C, Ti +2B), resulting in the synthesis of compounds.

There is a substantial literature that treats analytical aspects of the
initial--boundary value problem for different sharp-interface models with
kinetics \ related to the problem (\ref{he}-\ref{jc}) below, see \cite{luc,
vi, radkevich, chen, yin, yin1}. These works are concerned with basic issues
of mostly local in time existence. See also \cite{gz,xie} where a \textit{%
finite interval }version of the problem with linear kinetics and the
asymptotic behavior when a kinetic parameter tends to zero are discussed.
Dynamics on an semi-infinite spatial interrval for a one-phase model are
investigated in \cite{dhox}. The principal result of this paper is
asymptotics for the position of the front $s(t)$ of the form $s(t)\sim
kt^{\alpha }+O(t^{\beta })$ for $t\rightarrow \infty ,$ $0<\beta <\alpha $,
where $\alpha =1/2$ or $1$, depending on the value of undercooling. Needless
to say that these asymptotics, being important on their own right, cannot
capture order one variations in the temperature profile and velocity. These
variations, their compact structure and presumably low Hausdorff dimension
are the subject of our work (the graph of the time history of solutions in
Fig.~1 below gives some idea about their complexity). We also note recent
papers by Brauner \textit{et al., }\cite{brauner}-\cite{brauner1}, which
study dynamical behavior of solutions of a related problem. In particular
they consider perturbations of traveling-wave initial data and investigate
their instability and bifurcations.

The objective of our work is to investigate asymptotic behavior of a
propagating front. A variety of complex asymptotic dynamics (cf.~\cite
{ctm-port2}) arise only in the context of a infinite spatial interval. This
necessarily calls for a problem on an \textit{infinite }spatial domain, note
that for a bounded domain the flame front reaches the external boundary in
finite time and extinguishes. It is important to realize that the image of a
ball under the evolution is not compact for any finite time\textit{. }The
potential theory computations in Sec.\textit{~}3-4 allow us to extract the
part of the evolution that compactifies (the contribution from the free
interface), while the heat losses force the contribution from the initial
conditions decay with time exponentially. We believe that this clear
structure of the attractor indicates that careful potential theory estimates
are unavoidable if one is to study asymptotic patterns and prove existence
of a compact attractor in our case.

This study is partially motivated by numerical simulations and in particular
by the numerical experiments described in \cite{ctm-port2}, where it was
demonstrated that the system generates a remarkable variety of complex
thermokinetic oscillations. The dynamical patterns exhibited by the system,
as the governing parameters are varied, include a Hopf bifurcation (see the
rigorous proof \cite{vr}), period doubling cascades leading to chaotic
pulsations, a Shilnikov-Hopf bifurcation etc. These patterns are well-known
for the finite-dimensional dynamical systems and hint at the possibility
that the essential dynamics of the free-interface problem may be
finite-dimensional as well.

At the same time, in \cite{physd} we demonstrated that a $3\times 3$\ system
of ODEs obtained as a pseudo-spectral approximation to the \textit{one-phase}
free-boundary problem exhibits dynamics that mimics that of the
infinite-dimensional system to a surprising degree. For the one-phase
problem we \ were able to prove that compactness and finite dimensionality
of the attractor\ do take place \cite{jdde}.

It should be noted that the \textit{two-phase} problem is somewhat more
physically sound than its one-phase counterpart and appears in various
applications. However, the methods of the papers dealing with the one-phase
problem are not directly applicable to the two-phase Stefan problem with
kinetics which is the subject of the present communication. This is due to
the additional \ temperature field behind the propagating interface (in the
product phase) being not easily controllable. We overcome this difficulty in
the present paper and show that the temperature behind the interface is
sufficiently well-behaved to render compactness of the attractor.

Having proved compactness one is naturally curious as to how ''large'' is
the attractor in terms of some appropriate measure. Currently we are not
able to answer this question analytically due to, as we believe, purely
technical difficulties although we have little doubt that the dimension is
finite for the two-phase case. As the measure of complexity of the attractor
and therefore of the asymptotic regime, we use the correlation dimension
introduced by Grassberger and Procaccia (see, for example, \cite{grapro}).
The correlation dimension is based on the idea that if the evolution can be
described by a finite number of degrees of freedom then the time series of
observations on the system should be spatially correlated. This spatial
correlation is measured by the corellation sum, which is directly related to
the integral of the standard pair correlation function $c(\mathbf{r})$: 
\begin{equation*}
C(l)=\int_{\mathbf{|r|}\leq \,l}c(\mathbf{r})d\mathbf{r}
\end{equation*}
If \ the dimension is $\nu $ then it is easy to see that $C(l)\sim l^{\nu }$
for small $l$. We compute the correlation dimension of the attractors
generated in the direct numerical simulation of the problem.

The rest of the paper is organized as follows. In Sec.~2 we state the
free-boundary problem and present some minimal background information on
local existence and uniqueness. Sec.~3 and 4 constitute an analytical core
of the paper. Here we employ the single-layer representation for the
solution to obtain a natural decomposition of the solution into two
contributions, one from the initial conditions, and another one from the
free boundary. We obtain certain potential theory based estimates for both
contributions to the solution and its spatial derivative, which are
instrumental for the proof of compactness in Sec.~5. The estimates are
proved to be uniform with respect to the sup norm of initial data, assuming
that the nonlinearity in the kinetic free-boundary condition satisfies some
natural requirements.

In Sec.~5 we utilize the estimates that show that the contributions from the
free boundary are uniformly bounded and decay at infinity. Together with the
uniform bound on the spatial derivative, they allow us to apply a version of
the Arzela-Ascoli theorem, which guarantees that the contributions from the
free boundary for initial data from a fixed ball form a precompact set. We
complete the proof of existence of a compact attractor by using an
appropriate abstract result from dynamical systems. Finally in Sec.~6, from
direct numerical simulations of the free-interface problem we estimate the
correlation dimension of the attractor.

\section{The free-boundary problem. Local existence and uniqueness.}

We study the following free boundary problem: find $s(t)$ and $u(x,t)$ such
that 
\begin{equation}
u_{t}=u_{xx}-\gamma u,\quad x\neq s(t)  \label{he}
\end{equation}
\begin{equation}
u(x,0)=u_{0}(x)\geq 0  \label{ic}
\end{equation}
\begin{equation}
g[u(s(t),t)]=v(t)\quad \mathrm{for}\,t>0,  \label{kc}
\end{equation}
\begin{equation}
\lbrack u_{x}(s(t),t)]:=u_{x}^{+}(s(t),t)-u_{x}^{-}(s(t),t)=v(t)\quad 
\mathrm{for}\,\ t>0,  \label{jc}
\end{equation}
where $v(t)$ is the interface velocity, $s(t)=\int_{0}^{t}v(\tau )d\tau $ is
its position, $u$ is the temperature, and the derivatives $u_{x}^{+}$ and $%
u_{x}^{-}$ are taken from right side and left side of the free interface
respectively. The last term in the heat equation (\ref{he}) is due to the
heat losses into the medium surrounding the combustible or solidifying
substance via Newton's cooling law with a non-dimensional coefficient $%
\gamma \geq 0$.

The surrounding matter is assumed to be at the temperature of the fresh
combustible mixture at $-\infty $ (the original phase in the phase
transition interpretation). By the same token the heat loss will reduce the
temperature in the product phase to that of the medium. Thus the behavior of
the solution at infinity should satisfy $\lim\nolimits_{x\rightarrow \pm
\infty }u(x,t)=0$. It should be remarked that the presence of the heat
losses $\gamma >0\ $ only improves the analytical properties of the
solutions. For $\gamma =0$ the boundary condition at $\infty $ should be
replaced by convergence to a constant.

The dynamics of the physical system are determined by the feedback mechanism
between the heat release due to the kinetics $g(u|_{x=s(t)})$\ and the heat
dissipation by the medium. To illustrate the meaning of the kinetic term,
for example, in the context of solidification, we note that for some
substances in the presence of strong supercooling of the original phase the
phase transition temperature measured at the interface may deviate
considerably from the equilibrium one and is functionally related to the
interface velocity. This dependence called the interface attachment kinetics
can be different for different substances due to various microscopic
mechanisms responsible for the incorporation of the product at the interface
into the crystalline lattice.

The second interface condition (\ref{jc}) (the Stefan boundary condition)
expresses the balance between the heat produced at the free boundary and its
diffusion by the adjacent medium. As the problem describes propagation of
the phase transition front, the first interface condition (\ref{kc}) is a
manifestation of the \textit{nonequilibrium} nature of the transition; its
analog for the classical Stefan problem is just $u|_{x=s(t)}=0.$\ We should
mention that in contrast with the nonequilibrium problem, the dynamics of
the classical Stefan problem is relatively trivial.

To discuss properties of the attracting set for classical solutions of the
free interface problem (\ref{he})-(\ref{jc}) we need to first establish
their existence and uniqueness for all times. We say that $u(x,t),v(t)$ form
a \textit{classical solution} of (\ref{he})-(\ref{jc}) if

\begin{description}
\item  (i) $u(x,t)$ and $v(t)$ are continuous for $t\geq 0$;

\item  (ii) $u_{xx}$ and $u_{t}$ are continuous for $x\neq s(t),\ t>0$;

\item  (iii) Equations (\ref{he})-(\ref{jc}) are satisfied.
\end{description}

\noindent We quote the result as it was stated in \cite{minqu}:

\begin{theorem}
Suppose that the kinetic functions $g$ satisfies the following assumptions:%
\newline
$(A1)\ $ $g(u)$ is a continuously differentiable, monotone decreasing,
negative function on $(0,\infty )$ with $g(0)=-v_{0}$ for some velocity $%
-v_{0}<0;$\newline
$(A2)$ $\ g(u)$ is sublinear: ${\lim }_{u\rightarrow \infty }g(u)/u=0;$ 
\newline
and that the initial data $u_{0}(x)\geq 0$ \ are bounded. \newline
Then there exists one and only one classical solution $u(x,t)>0$ and $v(t)$
of the free interface problem (\ref{he})-(\ref{jc}). The solution is
uniformly bounded for all $t>0$.
\end{theorem}

For the reader's convenience we outline the scheme of the proof. First, the
problem is reduced to an integral equation for the interface velocity using
the single layer potential representation: 
\begin{equation}
u(x,t)=e^{-\gamma t}\int_{-\infty }^{\infty }G(x,t,\xi ,0)u_{0}(\xi )d\xi
-\int_{0}^{t}G(x,t,s(\tau ),\tau )e^{-\gamma (t-\tau )}v(\tau )d\tau ,
\label{iu}
\end{equation}
where 
\begin{equation}
G(x,t,\xi ,\tau )=\exp \{-\frac{(x-\xi )^{2}}{4(t-\tau )}\}[4\pi (t-\tau
)]^{-1/2}
\end{equation}
is the heat kernel and $s(t)=\int_{0}^{t}v(\tau )d\tau $.

Taking the limit of (\ref{iu}) as $x\rightarrow s(t)$ and using the kinetics
condition (\ref{kc}), we obtain an integral equation in terms of $v$ only:

\begin{equation}
g^{-1}(v(t))=e^{-\gamma t}\int_{-\infty }^{\infty }G(s(t),t,\xi ,0)u_{0}(\xi
)d\xi -\int_{0}^{t}G(s(t),t,s(\tau ),\tau )e^{-\gamma (t-\tau )}v(\tau
)d\tau ,  \label{ie}
\end{equation}

Next, we show that for sufficiently small time intervals, the mapping $K$
defined by the right hand side of the integral equation 
\begin{equation}
v(t)=g\left[ e^{-\gamma t}\int_{-\infty }^{\infty }G(s(t),t,\xi ,0)u_{0}(\xi
)d\xi -\int_{0}^{t}G(s(t),t,s(\tau ),\tau )e^{-\gamma (t-\tau )}v(\tau
)d\tau \right] :=Kv
\end{equation}
defines a contraction on an appropriately chosen closed set of continuous
functions which yields local existence. We also remark that the velocity $%
v(t)$ can be shown to be an infinitely differentiable function. The
centerpiece of the global existence proof is the \textit{a priori} estimate,
which allows us to extend the local solution indefinitely.

\section{A\textit{\ }priori estimates: spatial decay of solutions}

In order to demonstrate existence of an absorbing set we need to establish
spatial decay of the interface contribution to the classical solutions of
the problem. First we obtain an estimate for the solution on the interface.

\begin{theorem}
Let $u(x,t),v(t)$ be a classical solution of (\ref{he})-(\ref{jc}) and $%
||u_{0}||={{\sup }_{-\infty <x<\infty }}|u_{0}(x)|\leq M,$ then 
\begin{equation}
|u(s(t),t)|\leq 2M+R_{g},\qquad |v(t)|\leq g(2M+R_{g})
\end{equation}
where $R_{g}$ and $\Lambda $\ are constants dependent solely on the kinetic
function $g.$
\end{theorem}

\begin{proof}
First we prove the estimate for the interface temperature: $\psi
(t)=g(v(t))=I_{1}-I_{2},$ where $I_{1}$ and $I_{2}$ are the two parts of the
right hand side of (\ref{ie}).

It is rather obvious that $|I_{1}|\leq Me^{-\gamma t}$: 
\begin{equation*}
|e^{-\gamma t}\int_{-\infty }^{\infty }G(s(t),t,\xi ,0)u_{0}(\xi )d\xi |=%
\frac{e^{-\gamma t}}{2\sqrt{t\pi }}||u_{0}||\int_{-\infty }^{\infty }\exp \{-%
\frac{(s(t)-\xi )^{2}}{4t}\}d\xi =e^{-\gamma t}||u_{0}||
\end{equation*}

Now, since the kinetic function satisfies the condition (A2), for any $%
\varepsilon >0$ there exists $v_{1}>0$ such that $|g(\psi )/\psi |\leq
\varepsilon $ if $g(\psi )\leq -v_{1}.$ We subdivide $I_{2}$ as follows: 
\begin{equation}
I_{2}=\int_{0}^{t}G(s(t),t,s(\tau ),\tau )e^{-\gamma (t-\tau )}g(\psi (\tau
))d\tau =\int_{\chi _{1}}+\int_{\chi _{2}}=I_{3}+I_{4},
\end{equation}
where $\chi _{1}=\{\tau |-v_{1}<g(\psi (\tau ))<-v_{0},0<\tau <t\}\ $ and $%
\chi _{2}=(0,t)\setminus \chi _{1}$.

For $I_{3}\ $we have: 
\begin{gather*}
|\int_{\chi _{1}}G(s(t),t,s(\tau ),\tau )g(\psi (\tau ))d\tau |\leq
v_{1}|\int_{0}^{t}G(s(t),t,s(\tau ),\tau )d\tau | \\
=v_{1}|\int_{0}^{t}\exp \{-\frac{(s(t)-s(\tau ))^{2}}{4(t-\tau )}\}(2\sqrt{%
\pi (t-\tau )})^{-1}d\tau | \\
\leq v_{1}|\int_{0}^{\infty }-\frac{2}{v_{0}\sqrt{\pi }}\exp \{-v_{0}^{2}%
\frac{t-\tau }{4}\}d(\frac{v_{0}\sqrt{t-\tau }}{2})|=\frac{v_{1}}{v_{0}}.
\end{gather*}
Here we have used the observation that $|s(t)-s(\tau )|=|v(\xi )|(t-\tau
)\geq v_{0}(t-\tau )$ for some $\tau \leq \xi \leq t$.

Now, let us interpret $I_{4}(t)=P\psi (t)$ as a mapping; then the following
estimate holds for its norm: 
\begin{eqnarray}
\left\| P\psi \right\| &\leq &|\int_{\chi _{2}}G(s(t),t,s(\tau ),\tau
)g(\psi (\tau ))d\tau |  \notag \\
&\leq &|\int_{\chi _{2}}G(s(t),t,s(\tau ),\tau )\varepsilon \psi (\tau
)d\tau |  \notag \\
&\leq &\varepsilon \left\| \psi \right\| |\int_{0}^{\infty }G(s(t),t,s(\tau
),\tau )d\tau |\leq \frac{\varepsilon }{v_{0}}\left\| \psi \right\| .
\end{eqnarray}
Since $\psi =I_{1}-I_{3}-P\psi ,$ we have $\left\| \psi +P\psi \right\|
=||I_{1}(t)-I_{3}(t)||\leq M+v_{1}/v_{0}.$ On the other hand, by choosing $%
\varepsilon =v_{0}/2,$ we have $\left\| P\right\| \leq \frac{1}{2},$ and
therefore 
\begin{equation*}
|\psi (t)|=u(s(t),t)\leq 2M+2v_{1}/v_{0}.
\end{equation*}
The constant 
\begin{equation}
R_{g}:=2v_{1}/v_{0}  \label{Rg}
\end{equation}
is the constant referred to in the statement of the theorem$.$
Simultaneously, 
\begin{equation}
|v(t)|\leq g(2M+R_{g})  \label{v-bound}
\end{equation}
\end{proof}

Now using the uniform bounds for the interface velocity and temperature that
we have just established it is easy to obtain the uniform estimate for the
entire field through the maximum principle: 
\begin{equation}
|u(x,t)|\leq 2M+R_{g},  \label{uniform-m}
\end{equation}

From now on we shall assume that $g(u)$ is a monotonically decreasing
differentiable function on $[0,\infty ]$ with $|g^{\prime }|\leq C$ (recall
that the velocity $v=g(u)$ is negative) and 
\begin{equation}
v_{0}\leq -g(u)\leq V_{0}\;\mathrm{{for}\;{some}\;}V_{0},v_{0}>0.
\end{equation}
Both conditions are satisfied for the standard Arrhenius kinetics that in
appropriate rescaled and normalized variables has the form (\ref{arr}). The
existence of the lower bound $v_{0}$ in particular seems to be crucial for
the uniform boundedness of solutions.

For the compacness result we need certain decay estimates for the
contribution from the free boundary (see (\ref{front})-(\ref{behind}) below)
which are established in the the following

\begin{lemma}
\label{apriori2}Let $u(x,t),v(t)$ be a classical solution of (\ref{he})-(\ref
{jc}) with $||u_{0}||\leq M$ then for the contribution from the free
interface 
\begin{equation*}
\Psi (x,t)=\int_{0}^{t}G(x,s(\tau ),t-\tau )e^{-\gamma (t-\tau )}v(\tau
)d\tau
\end{equation*}
the following estimates hold: \newline
(i) 
\begin{equation}
|\Psi (x,t)|\leq \frac{V_{0}}{2\sqrt{\gamma }}  \label{uniform-gamma}
\end{equation}
(ii) For $x<s(t)$ 
\begin{equation}
|\Psi (x,t)|\leq \frac{V_{0}}{\sqrt{v_{0}^{2}+4\gamma }}\exp \{-\frac{%
v_{0}|x-s(t)|}{2}-\frac{|x-s(t)|^{2}}{4t}\}  \label{front}
\end{equation}
(iii) For $x>s(t)$ 
\begin{equation}
|\Psi (x,t)|\leq \left\{ 
\begin{array}{c}
\dfrac{V_{0}}{\sqrt{\gamma }}\exp (-\alpha (x-s(t)),\text{ for }%
x-s(t)>2V_{0}/\gamma \\ 
\dfrac{V_{0}}{\sqrt{\gamma }},\text{ for }0<x-s(t)<2V_{0}/\gamma
\end{array}
\right.  \label{behind}
\end{equation}
where $\alpha =\min (v_{0}/4,\gamma /2V_{0}).$
\end{lemma}

\begin{proof}
First we obtain a very simple bound, which is valid for any $x,t$: 
\begin{eqnarray*}
&&|\int_{0}^{t}e^{-\gamma (t-\tau )}\frac{e^{-(x-s(\tau ))^{2}/4(t-\tau )}}{%
\sqrt{4\pi (t-\tau )}}v(\tau )d\tau | \\
&\leq &V_{0}\int_{0}^{t}e^{-\gamma (t-\tau )}\frac{e^{-(x-s(\tau
))^{2}/4(t-\tau )}}{\sqrt{4\pi (t-\tau )}}d\tau \leq V_{0}\int_{0}^{\infty
}e^{-\gamma s}\frac{ds}{\sqrt{4\pi s}}=\frac{V_{0}}{2\sqrt{\gamma }}
\end{eqnarray*}
Note that this estimate is very different from the one obtained through the
maximun principle, (\ref{uniform-m}): the dependence on the norm of the
initial conditions is absent.

Ahead of the interface $x<s(t)$ 
\begin{eqnarray*}
&&|\int_{0}^{t}e^{-\gamma (t-\tau )}\frac{e^{-(x-s(\tau ))^{2}/4(t-\tau )}}{%
\sqrt{4\pi (t-\tau )}}v(\tau )d\tau | \\
&\leq &V_{0}\int_{0}^{t}e^{-\gamma (t-\tau )}\exp \{-\frac{%
(x-s(t))^{2}+2(x-s(t))(s(t)-s(\tau ))+(s(t)-s(\tau ))^{2}}{4(t-\tau )}\}%
\frac{d\tau }{\sqrt{4\pi (t-\tau )}} \\
&\leq &\frac{V_{0}}{\sqrt{\pi }}\exp \{-\frac{v_{0}|x-s(t)|}{2}-\frac{%
|x-s(t)|^{2}}{4t}\}\times \\
&&\int_{0}^{t}\exp \{-\frac{(s(t)-s(\tau ))^{2}}{4(t-\tau )}-\gamma (t-\tau
)\}\frac{d\tau }{2\sqrt{(t-\tau )}} \\
&\leq &\frac{V_{0}}{\sqrt{\pi }}\exp \{-\frac{v_{0}|x-s(t)|}{2}-\frac{%
|x-s(t)|^{2}}{4t}\}\int_{0}^{t}\exp \{(-\frac{v_{0}^{2}}{4}-\gamma )(t-\tau
)\}\frac{d\tau }{2\sqrt{(t-\tau )}} \\
&\leq &\frac{V_{0}}{\sqrt{v_{0}^{2}+4\gamma }}\exp \{-\frac{v_{0}|x-s(t)|}{2}%
-\frac{|x-s(t)|^{2}}{4t}\}
\end{eqnarray*}

To estimate the free-interface contribution to the solution $|\Psi (x,t)|$ 
\textit{behind the interface} $x>s(t)$ we split the interval of integration
into two subsets: $\chi _{1}=\{\tau \in \lbrack 0,t]:s(\tau )<(s(t)+x)/2\}$
and its complement $\chi _{2}=\{\tau \in \lbrack 0,t]:s(\tau )>(s(t)+x)/2\}$%
. 
\begin{equation*}
\int_{0}^{t}G(x,t,s(\tau ),\tau )e^{-\gamma (t-\tau )}\,|v(\tau )|d\tau
=\int\limits_{\chi _{1}}+\int\limits_{\chi _{2}}=I_{1}+I_{2},
\end{equation*}
For the first integral\ we have 
\begin{eqnarray*}
I_{1} &=&\int\limits_{\chi _{1}}\frac{\exp [-(x-s(\tau ))^{2}\frac{1}{%
4(t-\tau )}]}{2\sqrt{\pi (t-\tau )}}e^{-\gamma (t-\tau )}\,|v(\tau )|d\tau \\
&\leq &\frac{V_{0}}{2\sqrt{\pi }}\int\limits_{\chi _{1}}(t-\tau )^{-1/2}\exp
[-(x-s(t))^{2}\frac{1}{16(t-\tau )}]e^{-\gamma (t-\tau )}d\tau \\
&\leq &\frac{V_{0}}{2\sqrt{\pi }}\int\limits_{\chi _{1}}(t-\tau )^{-1/2}\exp
[-(x-s(t))\frac{v_{0}}{8}]e^{-\gamma (t-\tau )}d\tau \\
&\leq &\frac{V_{0}}{2\sqrt{\pi }}\exp [-(x-s(t))\frac{v_{0}}{4}%
]\int\limits_{0}^{(x-s(t))/(2v_{0})}\eta ^{-1/2}e^{-\gamma \eta }d\eta \\
&=&\frac{V_{0}}{2\sqrt{\gamma }}\func{erf}\left( \sqrt{\gamma \frac{x-s(t)}{%
2v_{0}}}\right) \exp [-(x-s(t))\frac{v_{0}}{4}]\leq \frac{V_{0}}{2\sqrt{%
\gamma }}\exp [-(x-s(t))\frac{v_{0}}{4}]
\end{eqnarray*}
The following inequalities 
\begin{equation*}
(x-s(\tau ))^{2}\frac{1}{(t-\tau )}\leq (\frac{x-s(t)}{2})^{2}\frac{1}{%
(t-\tau )}\leq (\frac{x-s(t)}{2})^{2}\frac{2v_{0}}{x-s(t)}
\end{equation*}
have been used to replace the exponent in the Gaussian kernel, which gave
rise to the exponential decay factor.

For the integral $I_{2}$ we obtain 
\begin{gather}
I_{2}=\dint\limits_{\chi _{2}}\frac{e^{-\gamma (t-\tau )}}{2\sqrt{\pi
(t-\tau )}}\exp [-\dfrac{(x-s(\tau ))^{2}}{4(t-\tau )}]\,|v(\tau )|d\tau 
\notag \\
\leq V_{0}\int\limits_{(x-s(t))/(2V_{0})}^{\infty }\frac{1}{2\sqrt{\pi \eta }%
}e^{-\gamma \eta }d\eta =\frac{V_{0}}{\sqrt{\pi }}\int\limits_{\sqrt{%
(x-s(t))/(2V_{0})}}^{\infty }\exp (-\gamma \xi ^{2})d\xi  \notag \\
\leq \left\{ 
\begin{array}{l}
\dfrac{V_{0}}{2\sqrt{\gamma \pi }}\exp (-\gamma (x-s(t))/(2V_{0})),\text{
for }\gamma (x-s(t))/(2V_{0})>1 \\ 
\dfrac{V_{0}}{2\sqrt{\gamma }},\text{ for }0\leq \gamma (x-s(t))/(2V_{0})<1
\end{array}
\right.  \label{i2-function}
\end{gather}
The final inequalties in the above estimate are based on the following
elementary observations: if $a\sqrt{b}>1$ then 
\begin{equation*}
\int_{a}^{\infty }\exp (-b\eta ^{2})d\eta =\frac{1}{\sqrt{b}}\int_{a\sqrt{b}%
}^{\infty }\exp (-\eta ^{2})d\eta \leq \frac{1}{\sqrt{b}}\int_{a\sqrt{b}%
}^{\infty }\eta \exp (-b\eta ^{2}\}d\eta =\frac{1}{2\sqrt{b}}\exp (-ba^{2});
\end{equation*}
on the other hand 
\begin{equation*}
\int_{a}^{\infty }\exp (-b\eta ^{2})d\eta \leq \int_{0}^{\infty }\exp
(-b\eta ^{2})d\eta =\frac{\sqrt{\pi }}{2\sqrt{b}}
\end{equation*}

Thus for $x>s(t)$ we obtain 
\begin{equation}
|\Psi (x,t)|\leq \left\{ 
\begin{array}{c}
\dfrac{V_{0}}{2\sqrt{\gamma }}\exp [-(x-s(t))\dfrac{v_{0}}{4}]+ \\ 
\dfrac{V_{0}}{2\sqrt{\gamma }}\exp (-\gamma (x-s(t))/(2V_{0})),\text{ for }%
\gamma (x-s(t))/(2V_{0})>1 \\ 
\dfrac{V_{0}}{2\sqrt{\gamma }}\exp [-(x-s(t))\dfrac{v_{0}}{4}]+ \\ 
\dfrac{V_{0}}{2\sqrt{\gamma }}\text{ for }0<\gamma (x-s(t))/(2V_{0})<1
\end{array}
\right.  \label{beh1}
\end{equation}
\end{proof}

Obviously the direct contribution from the initial conditions is bounded by
the norm of the initial conditions: \ 
\begin{equation}
e^{-\gamma t}\int_{-\infty }^{\infty }G(x,t,\xi ,0)u_{0}(\xi )d\xi \leq 
\frac{e^{-\gamma t}}{2\sqrt{t\pi }}||u_{0}||\int_{-\infty }^{\infty }\exp \{-%
\frac{(x-\xi )^{2}}{4t}\}d\xi =e^{-\gamma t}||u_{0}||  \label{i-data}
\end{equation}

\section{Estimate for the derivative}

The proof of compactness is based on a version of Arcela-Ascoli theorem and
uses an estimate for the derivative of the solution. Via differentiation of
the representation of the solution (\ref{iu}), the derivative for $x\neq
s(t) $ is expressed as follows: 
\begin{equation}
u_{x}(x,t)=-e^{-\gamma t}\int_{-\infty }^{\infty }G_{\xi }(x,t,\xi
,0)u_{0}(\xi )d\xi +\int_{0}^{t}G_{\xi }(x,t,s(\tau ),\tau )e^{-\gamma
(t-\tau )}v(\tau )d\tau ,  \label{ux}
\end{equation}

\begin{lemma}
Let $v(t)$ be a continuous function on $[0,T],$ define the derivative of the
boundary contribution as 
\begin{equation}
\Phi (x,t)=\frac{\partial }{\partial x}\int_{0}^{t}G(x,s(\tau ),t-\tau
)e^{-\gamma (t-\tau )}v(\tau )d\tau
\end{equation}
Then for every $0<t\leq T$ $\ \ \ \ |\Phi (x,t)|\leq const$
\end{lemma}

\begin{proof}
\textit{The estimate ahead of the front, i.e. for} $x\leq s(t),$ is treated
as follows. In the estimates ahead of the interface we replace $\exp
(-\gamma (t-\tau ))$ by $1.$ Consider separately two cases: $|s(t)-x|>1$ and 
$|s(t)-x|\leq 1$. \ \ \ 

For the case $|s(t)-x|>1$%
\begin{gather}
|\Phi (x,t)|=|\int_{0}^{t}\frac{x-s(\tau )}{2(t-\tau )}\frac{e^{-(x-s(\tau
))^{2}/4(t-\tau )}}{\sqrt{4\pi (t-\tau )}}e^{-\gamma (t-\tau )}v(\tau )d\tau
|  \notag \\
=|\int_{0}^{t}\frac{(x-s(\tau ))^{2}}{2(t-\tau )(x-s(\tau ))}e^{-(x-s(\tau
))^{2}/8(t-\tau )}\times e^{-(x-s(\tau ))^{2}/8(t-\tau )}\frac{v(\tau )d\tau 
}{\sqrt{4\pi (t-\tau )}}|  \notag \\
\leq |\int_{0}^{t}\frac{4/e}{s(t)-s(\tau )}\times  \notag \\
\exp \{-\frac{(x-s(t))^{2}+2(x-s(t))(s(t)-s(\tau ))+(s(t)-s(\tau ))^{2}}{%
8(t-\tau )}\}\frac{v(\tau )d\tau }{\sqrt{\pi (t-\tau )}}|  \notag \\
\leq \frac{4V_{0}}{v_{0}e\sqrt{\pi }}\int_{0}^{t}(t-\tau
)^{-3/2}e^{-(x-s(t))^{2}/8(t-\tau )}e^{-v_{0}|x-s(t)|/4}e^{-v_{0}^{2}(t-\tau
)/8}d\tau  \notag \\
\leq \frac{4V_{0}e^{-v_{0}|x-s(t)|/4}}{e\sqrt{\pi }v_{0}|s(t)-x|}%
\int_{0}^{t}|s(t)-x|(t-\tau )^{-3/2}e^{-(x-s(t))^{2}/8(t-\tau )}d\tau  \notag
\\
\leq \frac{32\sqrt{2}V_{0}e^{-v_{0}|x-s(t)|/4}}{e\sqrt{\pi }v_{0}|s(t)-x|}%
\int_{0}^{\infty }e^{-\eta ^{2}}d\eta \leq \frac{16\sqrt{2}V_{0}}{ev_{0}}%
\frac{e^{-v_{0}|x-s(t)|/4}}{|s(t)-x|}  \label{farahead}
\end{gather}
In the last estimate we used the following simple observations: $\xi e^{-\xi
}\leq 1/e,$ for $\xi =\dfrac{(x-s(\tau ))^{2}}{8(t-\tau )}>0,$ $|s(\tau
)-x|>|s(t)-x|$, \ $|s(\tau )-x|>|s(t)-s(\tau )|>v_{0}|t-\tau |$ and
substitution $\eta =|s(t)-x|(t-\tau )^{-1/2}/\sqrt{8}$ to obtain the error
function integral.

For the less involved case $|s(t)-x|\leq 1$ we split the integral into two
parts 
\begin{gather}
|\Phi (x,t)|=|\int_{0}^{t}\frac{x-s(\tau )}{2(t-\tau )}\frac{e^{-(x-s(\tau
))^{2}/4(t-\tau )}}{\sqrt{4\pi (t-\tau )}}v(\tau )d\tau |  \notag \\
\leq |\int_{0}^{t}\frac{|x-s(t)|+|s(t)-s(\tau )|}{2(t-\tau )}\frac{%
e^{-(x-s(\tau ))^{2}/4(t-\tau )}}{\sqrt{4\pi (t-\tau )}}v(\tau )d\tau | 
\notag \\
\leq \frac{V_{0}}{\sqrt{\pi }}\int_{0}^{t}\frac{|s(t)-x|(t-\tau )^{-3/2}}{4}%
e^{-(x-s(t))^{2}/4(t-\tau )}d\tau +\frac{V_{0}^{2}}{4\sqrt{\pi }}\int_{0}^{t}%
\frac{e^{-(s(t)-s(\tau ))^{2}/4(t-\tau )}}{\sqrt{(t-\tau )}}d\tau  \notag \\
\leq \frac{V_{0}}{\sqrt{\pi }}\int_{0}^{\infty }e^{-\eta ^{2}}d\eta +\frac{%
V_{0}^{2}}{4\sqrt{\pi }}\int_{0}^{t}\frac{e^{-v_{0}^{2}(t-\tau )/4}}{\sqrt{%
(t-\tau )}}d\tau  \notag \\
\leq \frac{V_{0}}{2}+\frac{V_{0}^{2}}{v_{0}\sqrt{\pi }}\int_{0}^{\infty
}e^{-\xi ^{2}}d\xi =\frac{V_{0}}{2}(1+\frac{V_{0}}{v_{0}})  \label{nearahead}
\end{gather}
where $\eta =(x-s(t))(t-\tau )^{-1/2}/2$ and $\xi =v_{0}\sqrt{(t-\tau )}/2.$

Thus, collecting estimates in (\ref{farahead})-(\ref{nearahead}) we observe
that the derivative ahead of the interface $x<s(t)$ decays exponentially 
\begin{equation}
|\Phi (x,t)|\leq \left\{ 
\begin{array}{l}
\dfrac{16\sqrt{2}}{ev_{0}}e^{-v_{0}|x-s(t)|/4}V_{0},\;x<s(t)-1 \\ 
\dfrac{V_{0}}{2}(1+\dfrac{V_{0}}{v_{0}}),\;s(t)-1\leq x\leq s(t)
\end{array}
\right.  \label{expdec}
\end{equation}

The part of the \textit{estimate concerning the derivative behind the
interface }$x>s(t)$ causes some difficulties due to the fact that $x$ can be
close or even equal to $s(\tau );$ it is treated as follows. For our
purposes it suffices to prove the estimate for $t$ starting from a certain $%
t>0$. Assume for convenience that $t>1$. For the second integral in (\ref{ux}%
) we split the interval of integration into two subsets: $\chi _{1}=\{\tau
\in \lbrack 0,t]:|s(\tau )-x|\geq |s(t)-x|/2\}$ and its compliment $\chi
_{2}=[0,t]\setminus \chi _{1}$. Note again that $s(t)$ is a monotone
function. We have 
\begin{gather}
\left| \int_{0}^{t}G_{\xi }(x,t,s(\tau ),\tau )e^{-\gamma (t-\tau )}v(\tau
)d\tau \right| \leq \int_{0}^{t}|G_{\xi }(x,t,s(\tau ),\tau )e^{-\gamma
(t-\tau )}|\times |v(\tau )|d\tau \\
=\int\limits_{\chi _{1}}+\int\limits_{\chi _{2}}=I_{1}+I_{2},
\end{gather}
For $I_{1}$ we use two subsets of $\chi _{1}$: $\chi _{1}=\chi _{11}\cup
\chi _{12}$ where $x>s(\tau )$ for $\chi _{11}$ and $x<s(\tau )$ for $\chi
_{12}$. For the integrals we get respectively 
\begin{eqnarray}
I_{11} &=&\int\limits_{\chi _{11}}\dfrac{|x-s(\tau )|}{2(t-\tau )}\times 
\frac{\exp [-\dfrac{(x-s(\tau ))^{2}}{4(t-\tau )}]}{2\sqrt{\pi (t-\tau )}}%
e^{-\gamma (t-\tau )}\times |v(\tau )|d\tau  \notag \\
&\leq &\frac{2V_{0}}{\sqrt{\pi }}\int\limits_{0}^{\infty }|x-s(t)|(t-\tau
)^{-3/2}\exp [-(x-s(t))^{2}\frac{1}{16(t-\tau )}]d\tau  \notag \\
&\leq &\frac{2V_{0}}{\sqrt{\pi }}\int\limits_{0}^{\infty }e^{-\xi ^{2}}d\xi
=V_{0}  \label{i11}
\end{eqnarray}
(note that on $\chi _{11}$ $|x-s(t)|/2<|x-s(\tau )|<|x-s(t)|).$ In the above
integral we have made a substitution $|x-s(t)|(t-\tau )^{-1/2}/4=\xi $. For
the second part 
\begin{eqnarray}
I_{12} &=&\int\limits_{\chi _{12}}\dfrac{|x-s(\tau )|}{2(t-\tau )}\times 
\frac{\exp [-\dfrac{(x-s(\tau ))^{2}}{4(t-\tau )}]}{2\sqrt{\pi (t-\tau )}}%
e^{-\gamma (t-\tau )}\times |v(\tau )|d\tau  \notag \\
&\leq &\int\limits_{\chi _{12}}\dfrac{|s(t)-s(\tau )|}{2(t-\tau )}\times 
\frac{\exp [-\dfrac{(x-s(\tau ))^{2}}{4(t-\tau )}]}{2\sqrt{\pi (t-\tau )}}%
e^{-\gamma (t-\tau )}\times |v(\tau )|d\tau  \notag \\
&\leq &\dfrac{V_{0}^{2}}{4}\int\limits_{0}^{t}\frac{e^{-\gamma (t-\tau )}}{%
\sqrt{\pi (t-\tau )}}d\tau \leq \dfrac{V_{0}^{2}}{4\sqrt{\pi \gamma }}%
\int\limits_{0}^{\infty }\frac{e^{-\tau }}{\sqrt{\tau }}d\tau =\dfrac{%
V_{0}^{2}}{4\sqrt{\gamma }}  \label{i12}
\end{eqnarray}
We remark that a time independent estimate for the above integral is also
valid.

For $I_{2}$ we shall replace a function of the type\ $xe^{-x^{2}}$ by its
maximum $1/\sqrt{2e}$ 
\begin{eqnarray}
I_{2} &=&\int\limits_{\chi _{2}}\dfrac{|x-s(\tau )|}{2(t-\tau )}\times \frac{%
\exp [-\dfrac{(x-s(\tau ))^{2}}{4(t-\tau )}]}{2\sqrt{\pi (t-\tau )}}%
e^{-\gamma (t-\tau )}\times |v(\tau )|d\tau  \notag \\
&\leq &V_{0}\int\limits_{\chi _{2}}\dfrac{|x-s(\tau )|}{2\sqrt{(t-\tau )}}%
\times \exp [-\dfrac{(x-s(\tau ))^{2}}{4(t-\tau )}]\times \frac{1}{2\sqrt{%
\pi }(t-\tau )}\,d\tau  \notag \\
&\leq &\dfrac{1}{\sqrt{8\pi e}}V_{0}\int\limits_{\chi _{2}}\frac{d\tau }{%
(t-\tau )}\leq \dfrac{1}{\sqrt{8\pi e}}V_{0}^{2}\int\limits_{\chi _{2}}\frac{%
d\tau }{|s(t)-x|/2}\leq \dfrac{V_{0}^{2}}{v_{0}\sqrt{2\pi e}}  \label{i2}
\end{eqnarray}
We have also replaced $(t-\tau )$ in the denominator by its minimum $%
|s(t)-x|/2\left\| v\right\| $ on $\chi _{2}$and observe that $meas(\chi
_{2})\leq |s(t)-x|/v_{0}$.

By collecting the estimates in (\ref{i11})-(\ref{i2}) we obtain a uniform
bound for the derivative behind the interface $x>s(t)$, 
\begin{equation}
|\Phi (x,t)|\leq V_{0}+\dfrac{V_{0}^{2}}{4\sqrt{\gamma }}+\dfrac{V_{0}^{2}}{%
v_{0}\sqrt{2\pi e}}  \label{der-behind}
\end{equation}
This concludes the proof of the lemma.
\end{proof}

We also note that the first term in (\ref{ux}) can be easily estimated, 
\begin{gather*}
\left| \int_{-\infty }^{\infty }G_{\xi }(x,t,\xi ,0)u_{0}(\xi )e^{-\gamma
t}d\xi \right| \leq e^{-\gamma t}\int_{-\infty }^{\infty }\dfrac{|x-\xi |}{2t%
}\times \frac{\exp [-\dfrac{(x-\xi )^{2}}{4t}]}{2\sqrt{\pi t}}|u_{0}(\xi
)|d\xi \\
\leq ||u_{0}||\frac{e^{-\gamma t}}{\sqrt{\pi t}}\int_{-\infty }^{\infty }%
\dfrac{|x-\xi |}{2\sqrt{t}}\times \exp [-\dfrac{(x-\xi )^{2}}{4t}]\times 
\frac{d\xi }{2\sqrt{t}}\leq ||u_{0}||\frac{e^{-\gamma t}}{\sqrt{\pi t}}
\end{gather*}
since the integral on the last line is equal to $2\int_{0}^{\infty }\eta
\exp (-\eta ^{2})d\eta =1,$ with $\eta =|x-\xi |/2\sqrt{t}.$ Thus we get the
following

\begin{corollary}
For any $t\geq t_{0},$ $t_{0}>0,$ the derivative of the solution is
uniformly bounded: 
\begin{equation}
|u_{x}(x,t)|=||u_{0}||\frac{e^{-\gamma t}}{\sqrt{\pi t}}+C
\end{equation}
\end{corollary}

We reiterate that the proof of compactness in the next section is based on a
version of Arcela-Ascoli theorem and uses the estimate for the derivative of
the free-interface contribution, $|\Phi (x,t)|\leq const$. Thus the
estimates of this section are crucial for the compactness result.

\section{Absorbing set and attractor\label{absorb-sec}}

\setcounter{equation}{0}

In this section we use the estimates obtained above to establish existence
of a bounded absorbing set and of the attractor which is compact in the
space of continuous functions. It can be easily verified that all the
estimates and analytical properties of the solutions can be obtained without
the heat losses. On the other hand the problem with the heat losses exhibits
uniform exponential decay in time of the contribution from the initial data
which is utilized in the proof of compactness of the attractor.

The integral representation (\ref{iu}) describes the evolution of the
initial temperature distribution $u_{0}$: $u(t)=T(t)u_{0}$. We think of the
evolution as taking place for the functions on $(-\infty ,\infty )$ in the
moving coordinate system attached to the free boundary $x^{\prime }=x-s(t)$.
Note that all the results in Sec.~3-4 are in tems of $x^{\prime }.$

It is convenient to split the semigroup operator $T$ into two parts: the
contribution of the free boundary 
\begin{equation}
T_{1}(t)u_{0}(x^{\prime })=-\int_{0}^{t}e^{-\gamma (t-\tau )}G(x^{\prime
}+s(t),t,s(\tau ),\tau )v(\tau )d\tau  \label{t1}
\end{equation}
and that of the initial data 
\begin{equation}
T_{2}(t)u_{0}(x^{\prime })=e^{-\gamma t}\int_{-\infty }^{\infty }G(x^{\prime
}+s(t),t,\xi ,0)u_{0}(\xi )d\xi  \label{t2}
\end{equation}

We rephrase the esimates in (\ref{i-data}) and (\ref{uniform-gamma}) as the
following result that establishes existence of an absorbing set for the
evolution.

\begin{proposition}
(i) The semigroup $T_{2}$ is uniformly exponentially contracting in $C$: 
\begin{equation*}
\sup_{u^{0}\in X}||T_{2}(t)u^{0}||\leq \exp (-\gamma t)N
\end{equation*}
for any ball 
\begin{equation*}
X=\{u\in C;\quad ||u||\leq N\}
\end{equation*}
(ii) For any $\varepsilon >0,$ the ball $B_{abs}:=\{u\in C:\quad |u|\leq
V_{0}/(2\sqrt{\gamma })+\varepsilon \}$ is an absorbing set for all bounded
subsets of $C$ for the semigroup $T.$ Here the radius of the absorbing ball
reflects the contribution of the free interface alone.\newline
\end{proposition}

Next we prove that the boundary contribution to the evolution, i.e. the
operators $T_{1}(t)$ are \textit{uniformly compact}. Namely, the following
proposition holds:

\begin{proposition}
For any $t_{0}>0$ the orbit of the ball $O(X,t_{0})=\cup _{t\geq
t_{0}}T_{1}(t)X$ is relatively compact in $C.$
\end{proposition}

\begin{proof}
For \ the version of Arzela-Ascoli theorem appropriate for $C$ it is
sufficient to have uniform boundedness for the derivative and uniform decay
of the family of functions as $|x^{\prime }|\rightarrow \infty .$ From Lemma%
\ref{apriori2} we see that for the initial data in a ball, the contribution
from the interface exhibits a spatial decay uniformly with respect to time.
On the other hand, the estimate (\ref{T1-derivative}) demonstrate that the
spatial derivative is uniformly bounded.

Now it is a simple matter to construct a finite $\varepsilon $-net for $%
O(X,t_{0}).$ First we choose a finite interval $-L\leq x^{\prime }\leq L,$
beyond which the functions of the family are smaller than $\varepsilon ,$ it
is possible to accomplish because of the uniform in time spatial decay. In
view of the uniform bound on the derivative, the functional family is
equicontinuous. Therefore the restricion of $O(X,t_{0})$ on to $[-L,L]$ is
compact by the regular Arzela-Ascoli theorem. By extending the elements of
the $\varepsilon $-net from $[-L,L]$ to the whole line by zero we obtain an $%
\varepsilon $-net \ in $O(X,t_{0})$.
\end{proof}

The properties of the evolution operator $T(t)$ described in the above
propositions allow us to apply the abstract general result (see, for
example, \cite[Chap. 1]{temam}) that in our situation can be stated as
follows:

\begin{theorem}
The $\omega $-limit set $\mathcal{A}$ of the absorbing set $B_{a}$ is a
global exponential compact attractor for the metric space $C$; $\mathcal{A}$
is the maximal attractor in $C$ and it is connected.
\end{theorem}

\section{Numerical dimension of attractor}

Having proved compactness one is naturally curious as to how ''large'' is
the attractor in terms of some appropriate measure? For the one-phase
problem we \ have been able to prove that the Hausdorff dimension of the
attractor\ is finite \cite{jdde}. However, due to the additional \
temperature field behind the propagating interface (in the product phase),
methods of the papers dealing with the one-phase problem are not directly
applicable to the two-phase problem. Currently we are not able to overcome
these, we believe, purely technical difficulties, although we have little
doubt that the dimension is finite for the two-phase case as well.

The question arises also, whether the presence of the temperature field
behind the front affects the ''size'' of the attractor in comparison with
the one-phase case (see \cite{wilmington}). We provide an answer to this
question by computing the correlational dimension of the attractors
generated via direct numerical simulation of the problem.

While the Hausdorff dimension is convenient for analytical estimates, it is
highly nontrivial to compute and requires too much storage and CPU time.
More convenient computationally is the \emph{correlation dimension. }%
Although in general $d_{corr}\leq d_{Hausdorff},$ they are usually very
close. We follow the now standard procedure for computation of the
correlation dimension \cite{grapro}. Namely, consider the set $%
\{U_{i},\;i=1,...N\}$ of points on the attractor $U_{i}=U(T+i\tau ),$ where $%
T$ $\gg 1$. We consider a discreet approximation of the solution in the
space $\mathbb{R}^{k}$ by sampling the solution at $k$ points $%
U_{i}=(u(x_{1},T+i\tau ),...,u(x_{k},T+i\tau ))$. We measure the spatial
correlation between the points on the discreet approximation of the
attractor with the correlation integral 
\begin{equation*}
C(l)=\lim_{N\rightarrow \infty }\frac{1}{N^{2}}\{\text{number of pairs with
the distance }\rho (U_{i},U_{j})<l\}
\end{equation*}
If for small $l,$ $C(l)$\ scales as $l^{\nu }$ then the correlation exponent 
$\nu $\ can be taken as the correlation dimension of the attractor $%
d_{corr}. $ For practical calculations the frequency of sampling $\tau $,
the number $k $ of points in space where the solution is sampled at each
time, and the number of samples $N$ are determined empirically. Similarly,
for the low sample dimension $k$\ a better approximation for $d_{corr}$ may
be obtained if the Euclidean distance $\rho $\ is modified by inclusion of a
weight.

To obtain a numerical approximation of the attractors we solve the initial
value problem (\ref{he})-(\ref{jc}) for sufficiently large time until the
asymptotic regime is attained. Obviously the dimension of the attractor
should not depend on the choice of initial data, which was confirmed by
direct numerical simulations. Problem (\ref{he})-(\ref{jc}) was solved in
the frame attached to the free boundary on a finite interval $\ [-L,L]$ with
the Dirichlet conditions $w(-L,t)=0$ simulating the decay of the solution at 
$-\infty ,$ and $\partial w(L,t)/\partial x=0$ corresponding to the
stabilization of the temperature in the product phase. According to our
observations the results are practically insensitive to the increase in the
interval length after $L\sim 30$ (see \cite{ctm-port2} for the details of
the numerical algorithm). We remark that in contrast with the one-phase case
where $L\sim 10$\ was sufficient, one needs a rather large spatial domain to
obtain consistent numerical results.

To represent different dynamical regimes we use the Arrhenius kinetics , 
\begin{equation}
V=g(u):=-\exp [\alpha \frac{u-1}{\sigma +(1-\sigma )u}],  \label{arr}
\end{equation}
where (in the context of combustion) $\alpha $ is proportional to the
activation energy (Zeldovich number), and $\sigma $ is the temperature ratio
of the fresh mixture and the product, see e.g. \cite{ctm-port2}.

Thus, the attractor is represented as a set in $\mathbb{R}^{k}$, where $k$
is the number of sampling points of temperature profiles. We choose time
snapshots of the solution for every $0.08$ in the interval of the asymptotic
regime ($200<t<1800$) and consider them as a discrete approximation of the
attractor in $\mathbb{R}^{k}$. The correlation dimension for this discrete
set is evaluated as explained above. As a control experiment we selected a
periodic asymptotic regime, $\alpha =4.5$. It is immediately confirmed that $%
d_{corr}\approx 1$ as one should expect.\FRAME{fhFU}{4.8845in}{4.6743in}{0pt%
}{\Qcb{Time history $0<t<4000$ for chaotic dynamics $u(x,t)$ vs. $x,t$.}}{}{%
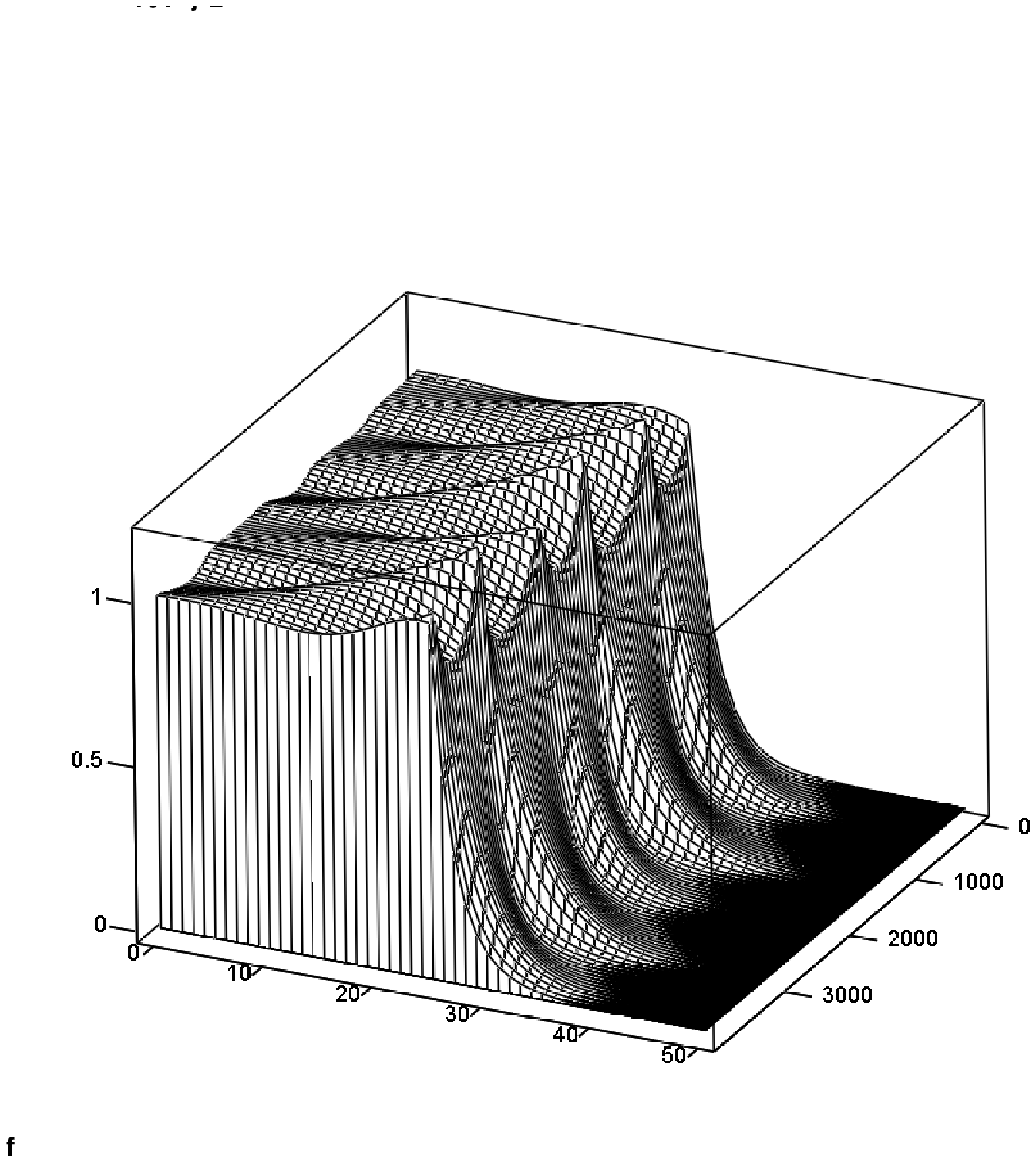}{\special{language "Scientific Word";type
"GRAPHIC";maintain-aspect-ratio TRUE;display "USEDEF";valid_file "F";width
4.8845in;height 4.6743in;depth 0pt;original-width 7.9693in;original-height
10.7695in;cropleft "0.1007";croptop "0.8042";cropright "0.7102";cropbottom
"0.3728";filename 'pictures/prof2p.ps';file-properties "XNPEU";}}

In contrast, for $\alpha =5,$ $\sigma =0.05$ the regime is chaotic as is
illustrated in Fig.~1 that presents a series of snapshots of spatial
temperature profiles. One can see from the log-log graph of the correlation
integral (Fig.~2) that in this case $d_{corr}\approx 2$. From our
observations on a variety of regimes it appears that the dimension cannot be
much higher than $2$.

\FRAME{fhFU}{2.1465in}{5.1171in}{0pt}{\Qcb{Correlation integral for 7-, 8-
and 9-point samples. }}{}{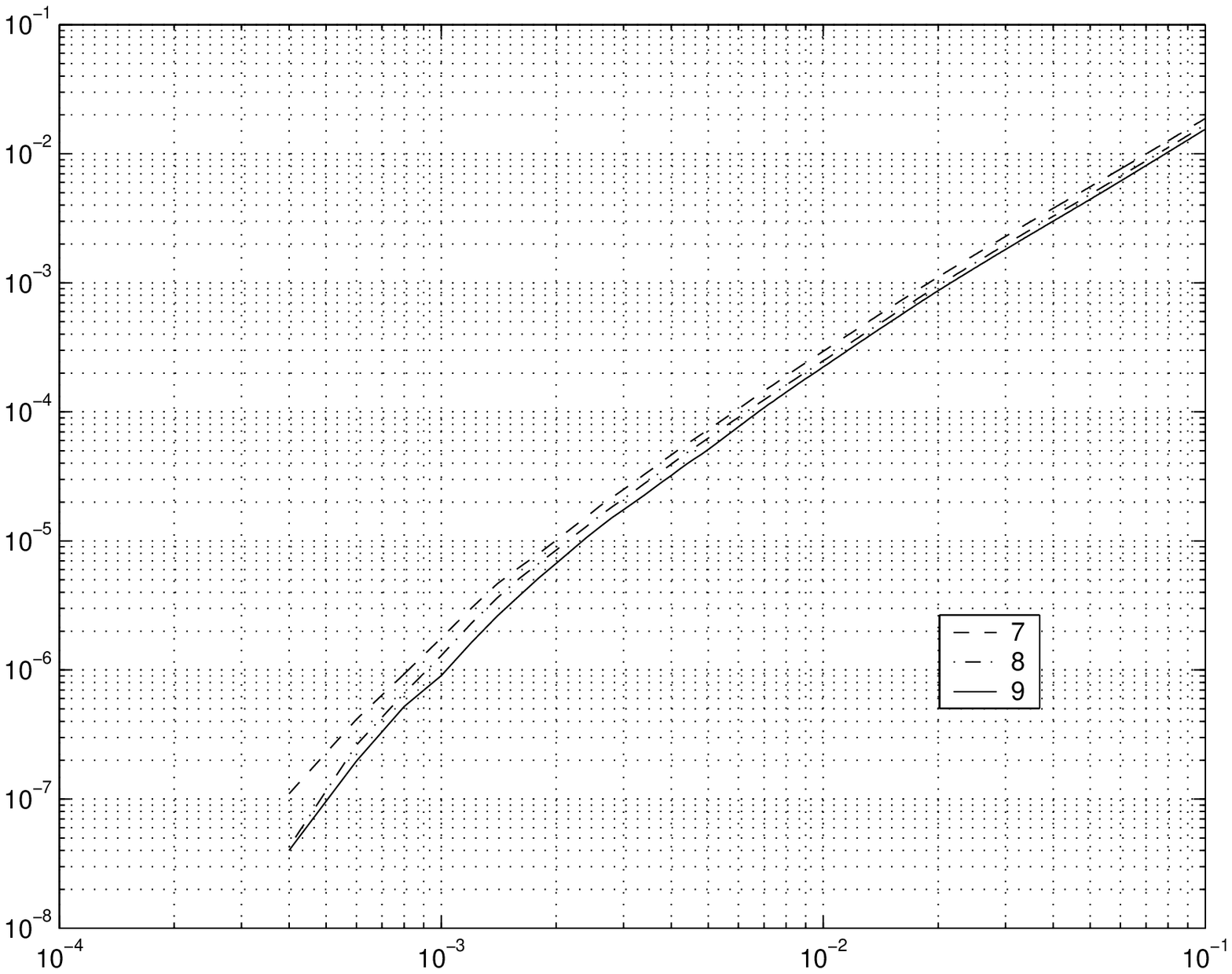}{\special{language "Scientific Word";type
"GRAPHIC";display "FULL";valid_file "F";width 2.1465in;height 5.1171in;depth
0pt;original-width 8.15in;original-height 10.6588in;cropleft "0";croptop
"1";cropright "1";cropbottom "0";filename
'pictures/fig2.eps';file-properties "XNPEU";}}

\section{Concluding remarks}

The compactness result has been proved here in the presence of heat losses
for any nonzero heat loss. Although we chose to operate in spaces of
continuous uniformly bounded functions on the infinite interval, we are
convinced that compactness can be established even for zero heat loss if
spaces with weaker topology are used, for instance in the space of
continuous functions bounded on each finite interval.

Results of this paper have been proved for the kinetic function satisfying
the bounds in (\ref{kinetics}). These bounds are quite physical and cover a
wide range of important applications. Nonetheless, our numerical
experimentation with different types of kinetic functions, including
unbounded ones demonstrate that the asymptotic dynamics are insensitive to
the behavior of the kinetic function for large temperatures. On the other
hand, our results from \cite{minqu} provide global existence and uniform
boundedness of solutions for a wider class of kinetic functions, namely for
sublinear kinetics. Therefore we strongly believe that the principal result
of this paper holds for this case as well.

It is interesting to compare the proof of the compactness above to that for
the one-phase problem \cite{ejde}. Although the estimates in the two-phase
case are more involved due to the presence of the temperature field behind
the propagating interface, once they are obtained, the representation of the
evolution semigroup is more transparent than in the one-phase case.

Also, it is rather remarkable that the numerical estimates for the
correlation dimension of the attractor above and for the one-phase case (see 
\cite{wilmington}) yield roughly the same value. Indeed, such an outcome is
rather unexpected because the two-phase problem seems to possess more
''degrees of freedom'' than its one-phase counterpart.

Finally, for the one-phase problem we \ have been able to prove that the
Hausdorff dimension of the attractor\ is finite \cite{jdde}. Currently we
are not able to overcome certain difficulties that we believe are of purely
technical nature but it appears safe to conjecture that the dimension of the
attractor is finite for the two-phase problem as well.

\section{Acknowledgments}

V. Roytburd is grateful to Gene Wayne for illuminating discussions. The
authors would like to acknowledge constructive and detailed suggestions by
the referees. This research was supported in part by NSF through grants
DMS-0207308 (MF) and DMS-9704325 (VR).

\end{document}